\documentclass[fleqn,10pt]{wlscirep}
\usepackage[utf8]{inputenc}
\usepackage[T1]{fontenc}
\usepackage{graphicx} 
\usepackage{braket}
\usepackage[title]{appendix}
\title{Realizing Topological Quantum Walks on NISQ Digital Quantum Computer}

\author[1,2,*]{Mrinal Kanti Giri}
\author[2]{Sudhindu Bikash Mandal}
\affil[1]{Department of Physics, National Tsing Hua University, Hsinchu 30013, Taiwan}
\affil[2]{Centre for Quantum Engineering, Research and Education, TCG CREST, Kolkata, India}
\affil[*]{mkg@mx.nthu.edu.tw}
\affil[*]{mrinalphy333@gmail.com}

\begin{abstract}
We study the quantum walk on the off-diagonal Aubry-Andr\'e-Harper (AAH) lattice with periodic modulation using a digital quantum computer. We investigate various initial states at the single-particle level, considering different hopping modulation strengths and phase factors. Initiating the quantum walk with a particle at the lattice edge reveals the robustness of the edge state, attributed to the topological nature of the AAH model, and displays the influence of the phase factor on this edge state. On the other hand, when the quantum walk begins with a particle in the lattice bulk, we observe a repulsion of the bulk walker from the edge, especially under strong hopping modulation. Furthermore, we extend our investigation to the quantum walk of two particles with nearest-neighbour (NN) interaction. We show the repulsion effect in the quantum walk when two walkers originate from the edge and bulk of the lattice due to the interaction. Additionally, when two particles are positioned at NN sites and subjected to strong hopping modulation strength, they exhibit localization in the presence of interaction. We analyze these phenomena by examining physical quantities such as density evolution, two-particle correlation, and participation entropy, and discuss their potential applications in quantum technologies.
\end{abstract}

\begin{document}

\maketitle

\section{Introduction}
The quantum walk holds great significance in the fields of quantum information science and fundamental physics~\cite{Nielsen,QWAharonov,Childs2002,Kempe2003,Venegas-Andraca2012,Manouchehri2014}. Its inherent quantum nature and ballistic behavior establish a versatile and potent foundation for various quantum information applications, such as solving search problems and enhancing the efficiency of quantum algorithms~\cite{Childs2004,ambainis2003quantum,Kempe2003,ambainis2007,Childs2014,Chakraborty2016}. Furthermore, quantum walks have potential applications in quantum computing~\cite{Childs_2009,Childs2013}, quantum metrology~\cite{metrology}, quantum biology~\cite{Mohseni2008,biological_quantum_walk_2014}, quantum simulation of physical phenomena, and quantum state engineering.
Quantum walks are generally two types: continuous-time and discrete-time walks. In this article, we will discuss the continuous-time quantum walk (QW), which provides a versatile approach to studying the dynamical properties of particles in a lattice. Recent experimental progress has facilitated the exploration and detection of QWs in diverse systems, including nuclear magnetic resonance (NMR)~\cite{PhysRevA.67.042316,PhysRevA.72.062317}, trapped atoms~\cite{Karski2009}, ion traps~\cite{trap1,trap2}, cold atoms in optical lattices~\cite{Karski2009,Greiner_walk}, optical waveguide arrays~\cite{silberberg2008,Peruzzo2010,poulios2014quantum}, and superconducting circuits~\cite{Yan2019,Ye2019}. These developments have resulted in significant theoretical investigations and experimental observations related to dynamic phase transitions, localization transitions, particle statistics, topological properties, Bloch oscillations, bound states, chiral dynamics, entanglement dynamics, and spin dynamics~\cite{Zhang2017,Li2023,Xie2019,Xie2020,Greiner_walk,Tai2017,bloch_magnon_expt,roos2019,gadway2018,Vovrosh2021,bloch_spin_charge_sep}.
In recent years, there has been a significant drive towards achieving fault-tolerant quantum computers, particularly using superconducting qubits and ion traps. However, current operational quantum computers, known as Noisy Intermediate Scale Quantum (NISQ) devices, face challenges like scalability, short decoherence times, lower gate fidelity and high error rates~\cite{Preskill2018}. Despite facing challenges, quantum computers demonstrate proficiency in efficiently simulating quantum systems, with Hamiltonian simulation emerging as a prominent application for NISQ computers. Within this framework, QWs have emerged as a crucial tool for simulating complex quantum systems and developing quantum algorithms.

In this work, we investigate the QW of particles on the commensurate off-diagonal Aubry-Andr\'e-Harper (AAH) model with NN interaction using the digital quantum computer provided by IBM Q. The AAH model has been theoretically extensively studied for its localization transition, multifractality, topological edge states and topological adiabatic pumping in both commensurate and incommensurate lattice scenarios~\cite{PhysRevB.28.4272,PhysRevLett.62.2714,PhysRevLett.83.3908,PhysRevB.87.134202,PhysRevB.55.12971,PhysRevLett.126.080602,sdas2013,PhysRevA.95.013619,PhysRevB.91.014108,Chaohong2016}. Beyond theoretical advancements, the AAH model has also been experimentally realized in various systems, such as cold atoms in optical lattices~\cite{Schreiber2015,Bordia2017}, photonic lattices~\cite{Chaohong2016,PhysRevLett.110.076403,PhysRevLett.109.106402}, and superconducting qubits~\cite{PhysRevLett.131.080401,Li2023}. 

We present a digital quantum simulation to explore the QW within the commensurate off-diagonal AAH model. We investigate various initial states at the single-particle level, considering different hopping modulation strengths and phase factors. When a particle initiates the QW from the lattice edge, we demonstrate the robustness of the edge state due to the topological nature of the AAH model and examine the influence of the phase factor on this edge state. Conversely, when a particle starts the QW from the lattice bulk, we observe repulsion from the edge, particularly with strong hopping modulation strength. We also investigate the QW of two particles with NN interaction. Although the non-interacting limit, particularly edge states and localization transitions within this model, has been experimentally realized using superconducting qubit processors~\cite{Li2023}, the effects of different initial states and particle interactions remain unexplored in the digital quantum computing. We show the repulsion effect in the QW when two walkers originate from the edge and bulk of the lattice due to the NN interaction. Additionally, when two particles are positioned at NN sites and subjected to strong hopping modulation strength, they exhibit localization in the presence of interaction. We study these properties by calculating density evolution, correlation, and participation entropy.

\section{Model and Approach}
We study the QW of spinless fermions or hardcore bosons on a one-dimensional lattice, focusing on the interacting off-diagonal AAH model with periodic hopping modulation. The corresponding Hamiltonian is given by -

\begin{equation}
    \hat{H} = J\sum_{i=1}^{N-1} [1 + \lambda_J \cos{(\frac{2\pi i}{T} + \phi_J)}] (\hat{c}^{\dagger}_{i+1} \hat{c}_{i} + \text{h.c.}) + V \sum_{i}^{N} \hat{n}_{i} \hat{n}_{i+1}. 
\label{eq:Ham_eq}
\end{equation}
Here, $\hat{c}^{\dagger}_{i}$ ($\hat{c}_{i}$) represents the fermionic creation (annihilation) operator at site $i$. $J$ represents nearest-neighbour hopping strength and $\hat{n}_i$ is the local number operator at the site $i$. The hopping term introduces inhomogeneity through a cosine modulation with strength $\lambda_J$. This cosine modulation has periodicity $T$ and phase factor $\phi_J$. $V$ denotes the nearest-neighbour interaction among particles.  Our analysis focuses on cases where $T$ is commensurate and an integer, specifically emphasizing on $T=2$ throughout the report. 
Our studies focused on the continuous-time quantum walk, following the unitary time evolution of the time independent Hamiltonian given in Eq.~\ref{eq:Ham_eq}, described as -
\begin{equation}
    |\psi(t)\rangle = e^{-i\hat{H}t} |\psi(0)\rangle,
\end{equation}
where $|\psi(0)\rangle$ representing some initial state. To comprehend the dynamics, we compute the local observable density evolution of the particles through -
\begin{equation}
    n_i(t) = \langle \psi(t) | \hat{n}_i |\psi(t)\rangle,
    \label{eq:density_op}
\end{equation}
where $\hat{n}_i$ denotes the density operator at lattice site $i$.
To simulate the Hamiltonian given in Eq.~\ref{eq:Ham_eq} on a quantum computer, we utilize the Jordan-Wigner transformation as detailed in Appendix-\ref{A:JW}. The resulting transformed Hamiltonian is given by-
\begin{equation}
    \hat{H} = \frac{J}{2}\sum_{i=1}^{N-1}  [1 + \lambda_J \cos{(\frac{2\pi i}{T} + \phi_J)}]\; (\hat{\sigma}_i^x\hat{\sigma}_{i+1}^x+ \hat{\sigma}_i^y\hat{\sigma}_{i+1}^y)
    + \frac{V}{2} \sum_{i}^{N} \hat{\sigma}_{i}^z \hat{\sigma}_{i+1}^z,
\label{eq:Ham}
\end{equation}
where $\hat{\sigma}_{i}^{k}$ represents Pauli matrices ($k=x,y,z$) with eigenvalues $\pm 1$. This transformed Hamiltonian is suitable for efficient simulation on the existing NISQ computers. 
We used qubits to represent lattice sites, where $\ket{0}$ corresponds to an unoccupied site and $\ket{1}$ to an occupied site. This maps directly to the Fock states of spinless fermions or hardcore bosons.
\begin{figure}[h!]
    \centering
    \includegraphics[width =0.9\columnwidth]{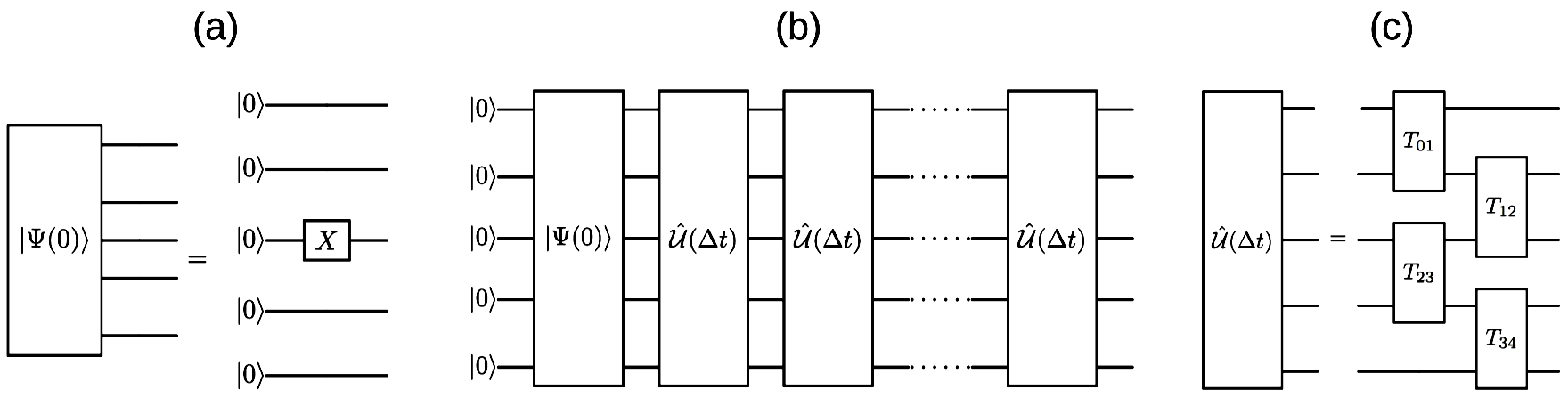}
    \caption{(a) The initial state $\ket{\psi(0)} = \ket{00100}$ in a 5-qubit system is prepared by applying NOT gate at third qubit. (b) The time evolution operator $\hat{U}$ is discretized into discrete steps and applied to the initial state (c) The discrete unitary operators $\hat{U}(\Delta t)$ are decomposed using basic trotterization.}
    \label{fig:circuit}
\end{figure}
Our simulation approach involves the following stages:
\begin{enumerate}
    \item \textbf{Initialization:} The initial state is generated by applying NOT gate(s) to the default state of the qubits on the IBM machine. For example, to create the state $\ket{\psi(0)} = \ket{00100}$ on a 5-qubit system, a NOT gate is applied to the third qubit, as shown in Fig.~\ref{fig:circuit}(a).

    \item\textbf{Time Discretization:} To implement the time evolution operator $\hat{U}$ on the initial state $\ket{\psi(0)}$ in the IBM machine, we discretized time ($t$), expressing the evolution operator as the product of discrete time steps ($\Delta t$) -
    \begin{equation}
        \hat{U}(t) = \hat{U}(\Delta t) \hat{U}(\Delta t) \dots \hat{U}(\Delta t).
    \end{equation}
    An associated diagram is provided in Fig.~\ref{fig:circuit}(b).

    \item\textbf{Trotterization:} The discrete unitary operators $\hat{U}(\Delta t)$ undergo trotterization into a product of one and two-qubit unitaries ($T_{ij}$). In this study, we have utilized a basic trotterization method, as detailed in Appendix-\ref{B:trotter}. The corresponding decomposition can be visualized in Fig.~\ref{fig:circuit}(c). 

    \item\textbf{Gate Decomposition:} The unitary ($T_{ij}$) is decomposed into CNOT and single-qubit rotation gates, suitable for direct application on IBM devices has been given in Ref.~\cite{PhysRevA.69.032315, PhysRevA.63.062309, Smith2019} and is shown in Fig.~\ref{fig:T-gate}.
    \begin{figure}[h!]
        \centering
        \includegraphics[width =0.8\columnwidth]{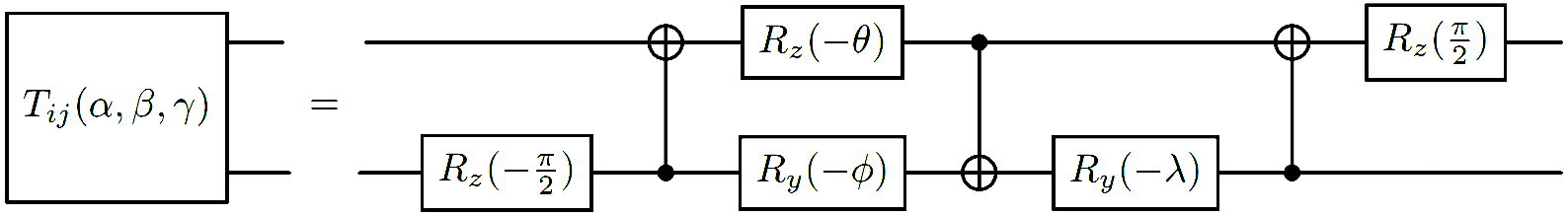}
        \caption{The quantum circuit corresponding to the unitary operator, $T_{ij}(\alpha,\beta,\gamma) = exp[-i(\alpha \hat{\sigma}^{x}_{i} \hat{\sigma}^{x}_{j} + \beta \hat{\sigma}^{y}_{i} \hat{\sigma}^{y}_{j} + \gamma \hat{\sigma}^{z}_{i} \hat{\sigma}^{z}_{j})]$. Here, $\theta = \frac{\pi}{2} - 2\gamma$, $\phi = 2\alpha - \frac{\pi}{2}$ and $\lambda = \frac{\pi}{2} - 2\beta$. }
        \label{fig:T-gate}
    \end{figure}
    
    \item\textbf{Measurement and Analysis:} Following the evolution process, the dynamical quantity of interest, measured in the computational basis as defined in Eq~\ref{eq:density_op}, is expressed as -
    \begin{equation}
        n_i(t) = \langle \psi(t) | \hat{\sigma}^z_i |\psi(t)\rangle.
    \end{equation}
\end{enumerate}
These steps collectively form the basis of our simulation approach, enabling the analysis of QW in the chosen Fock state representation.

\section{Quantum simulation on IBM quantum computers}

In this study, we present outcomes from simulations conducted on the 127-qubit IBM "ibm\_brisbane" instance. To benchmark our results, we compare them with those from an exact calculation ("Exact") and an ideal Qasm-simulator ("Q-sim") provided by Qiskit python API~\cite{qiskit}. To address potential errors from Hardware noise, we also employ readout error mitigation (REM)~\cite{Kandala2019, PhysRevLett.119.180509} using the Qiskit runtime (by setting optimization level 2) and Post-selection~\cite{Smith2019, PhysRevLett.122.180501}. It is crucial to acknowledge that IBM machines undergo re-calibration daily, introducing potential variations in data across different days. We examine some important analyses on the 7-qubit IBM instance "ibm\_oslo" device to address this. This approach aggregates data from different times and days, enhancing the robustness and inclusivity of our analysis. This approach ensures qualitative reproducibility and highlights the reliability of our results across different IBM machines. To overcome issues with reduced two-qubit gate fidelity and errors from prolonged execution times, we employed a linear topology and used basic trotterization for up to $10$ steps. Choosing a linear topology is crucial to ensure accurate simulation with superconducting qubits, aiming to avoid the requirement of SWAP operations.
\section{Results}
\begin{figure}[h!]
    \centering
    \includegraphics[width =0.75\columnwidth]{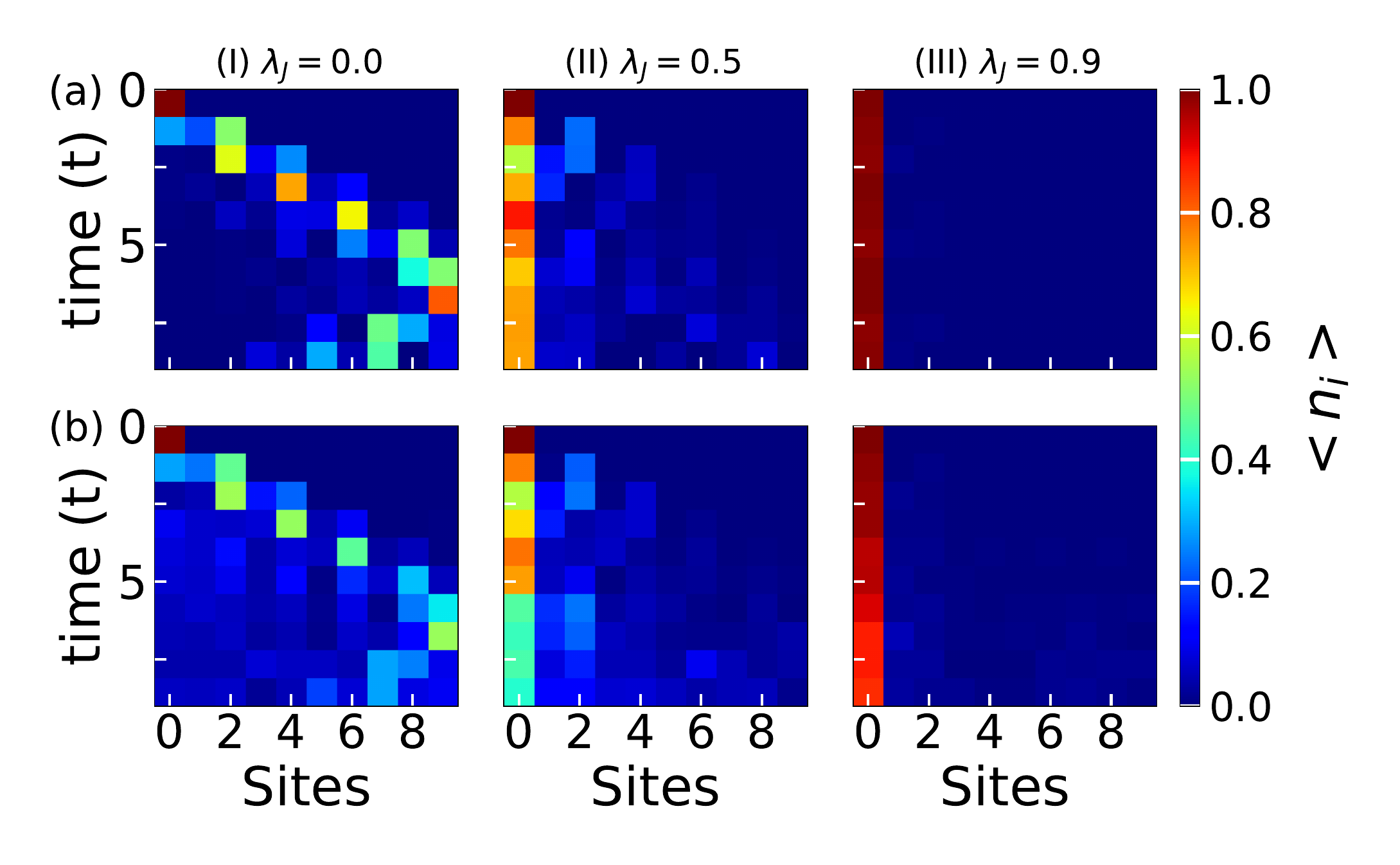}
    \caption{The figure depicts the density evolution of quantum walkers initialized with the state $\ket{\psi}_{I} = \hat{c}^\dagger_0 \ket{vac}$. Panels (a) and (b) showcase results from "Q-sim" and the readout error mitigated (REM) results from "ibm\_brisbane," respectively. Figures (I), (II), and (III) correspond to amplitude modulations of $\lambda_J=0.1$, $0.5$, and $0.9$, respectively. Here, we consider $\phi_J=0$ and $L=10$.}
    \label{fig:den_0}
\end{figure}
\subsection{Single-Particle QW} 
In this section, we study the single-particle QW within the framework of the off-diagonal AAH model with periodic modulation, characterized by the Hamiltonian defined in Eq.~\ref{eq:Ham}. We have considered three initial states with $L$ lattice sites, which are given by -
\begin{equation}
    \ket{\psi}_I = \hat{c}^\dagger_0 \ket{vac},
    \label{eq:psi_0}
\end{equation}
where the particle resides at the left edge,
\begin{equation}
    \ket{\psi}_{II} = \hat{c}^\dagger_{L-1} \ket{vac},
    \label{eq:psi_1}
\end{equation}
with the particle positioned at the right edge of the lattice and 
\begin{equation}
    \ket{\psi}_{III} = \hat{c}^\dagger_{L/2} \ket{vac},
    \label{eq:psi_2}
\end{equation}
where the particle is situated in the bulk of the system and $\ket{vac}$ is the empty state.

\begin{figure*}[h!]
  \centering
  \begin{minipage}[b]{0.49\columnwidth}
    \includegraphics[width=1\columnwidth]{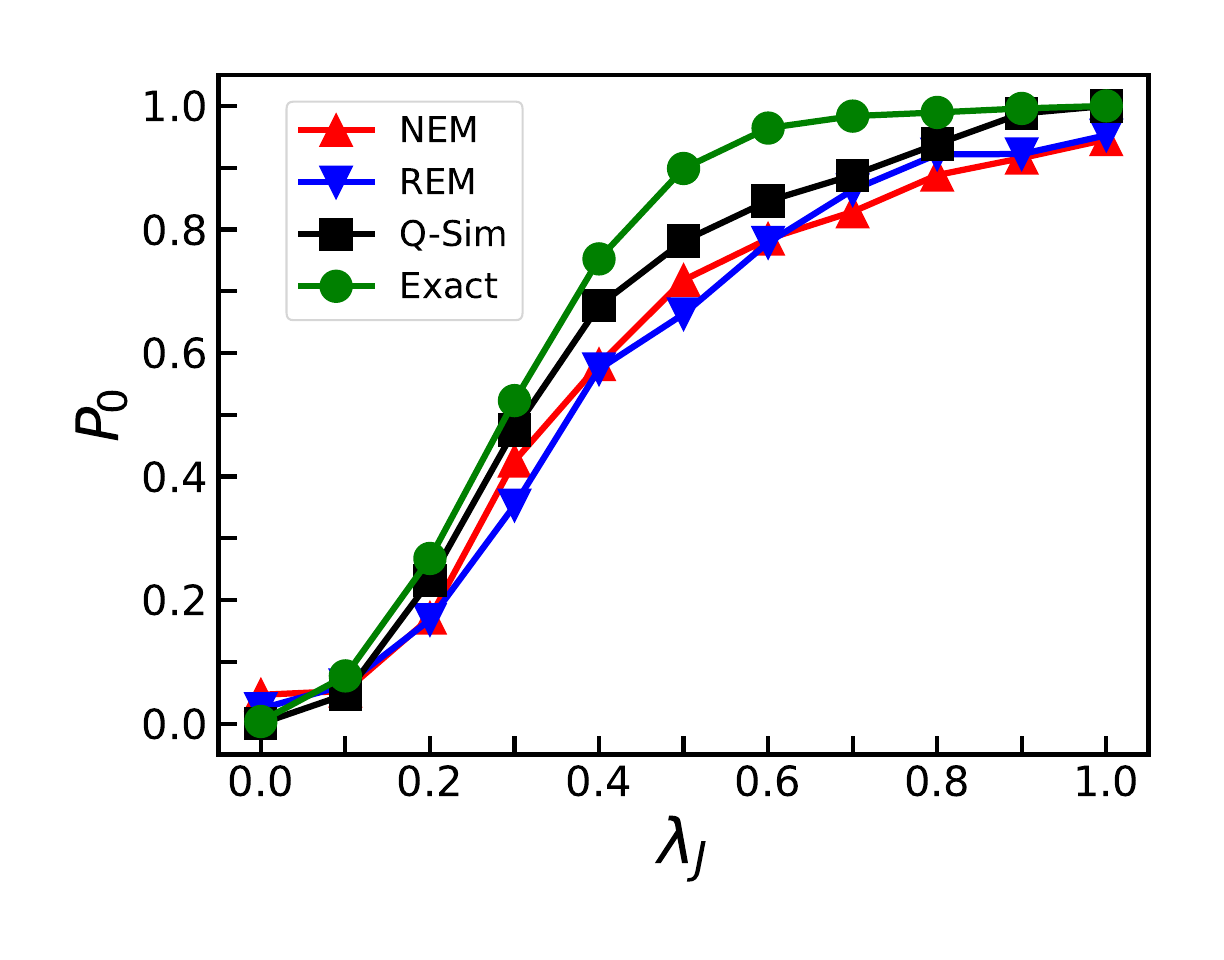}
    \caption{ The figure displays $P_{0}$ at time $t=5J^{-1}$ for various $\lambda_{J}$ values, considering $\phi_J=0$. The data includes both No Error Mitigation (NEM) and Readout Error Mitigation (REM) data obtained from the "ibm\_brisbane" machine.}
    \label{fig:critic}
  \end{minipage}
  \hfill
  \begin{minipage}[b]{0.49\columnwidth}
    \includegraphics[width=1\columnwidth]{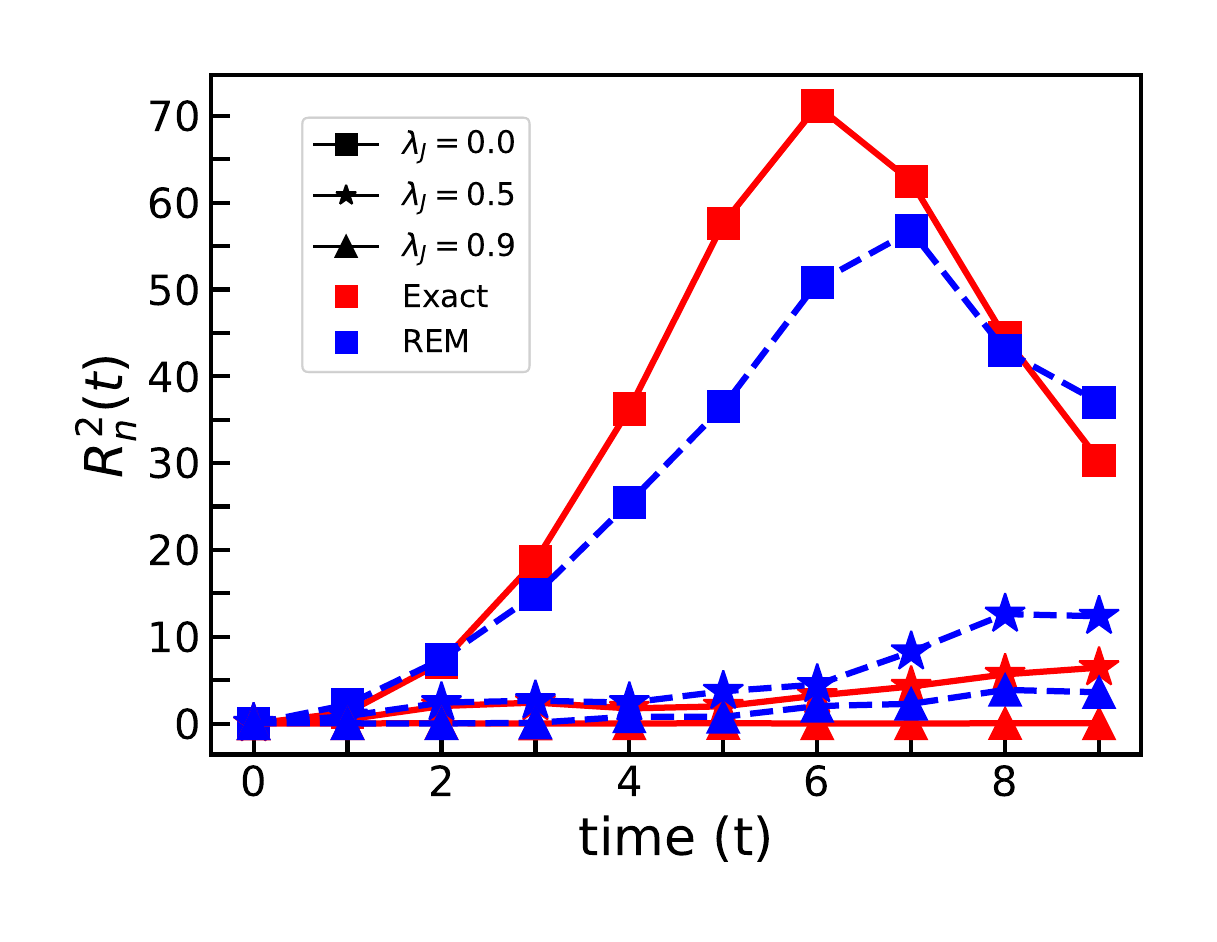}
    \caption{The figure displays $R_{n}^{2}(t)$ over time, t, for $\lambda_J = 0.0, 0.5$, and $0.9$. In this context, $\phi_J=0$. The data labeled REM corresponds to Readout Error Mitigation data obtained from the "ibm\_brisbane".}
    \label{fig:radial}
  \end{minipage}
\end{figure*}

We begin our investigation by setting the phase factor,$\phi_J=0$. Following this, we study the temporal evolution of the initial state $\ket{\psi}_{I}$ under the Hamiltonian, as specified in Equation~\ref{eq:Ham}. The dynamical evolutions are conducted both in the Q-sim and on the IBM machine, considering three distinct values of the hopping modulation strengths, namely $\lambda_J = 0.0, 0.5, \text{ and } 0.9$ and the corresponding results are depicted in Figure~\ref{fig:den_0}(I)-(III). 
For $\lambda_J=0.0$, the walker undergoes a unidirectional QW characterized by light-cone-like spreading reflects upon reaching the boundaries, as illustrated in Fig.~\ref{fig:den_0}a(I). As $\lambda_J$ increases, the walker gradually start to localize at the edge of the lattice. While in the case of large $\lambda_J$, complete localization is observed at the edge, as depicted in Fig.~\ref{fig:den_0}a(III).
This distinctive localization at the of the edge of the lattice is attributed to the emergence of a topologically protected edge state, discussed in references~\cite{sdas2013,PhysRevA.95.013619}. It is crucial to emphasize that analogous phenomena are evident in IBM machines, as exemplified in Panel-b of Fig.~\ref{fig:den_0}. Here, notable errors arise from trotter errors and inherent inaccuracies within the system after a few steps. 
It is important to note that, in accordance with the fundamental concepts of bulk-edge correspondence, this edge state consistently manifests across a range of periodic modulation strengths~\cite{sdas2013,PhysRevA.95.013619}. But, in the QW, localization behaviour is not prominent for weak modulation strengths.
To quantify this, we have plotted the density at the first site, $P_{0}=\langle \hat{n}_{0}\rangle$ as a function of the modulation strengths, $\lambda_J$ at time, $t=5J^{-1}$, as shown in Fig.~\ref{fig:critic}. It clearly shows the edge localization gradual increase with an increase in hopping modulation strength. Notably, IBM machine results align well with exact results, especially for larger $\lambda_J$ values. This modulation in QWs highlights its potential for studying the dynamics of quantum information, allowing control over the information spread and speed through the hopping amplitude parameter.
In Fig.~\ref{fig:den_0}, the density of the time-evolved wavefunction on the IBM machine deviates significantly from the exact results. This discrepancy is primarily due to the use of basic trotterization with a large step size ($\Delta t = 1$) in the time evolution. Interestingly, as the value of $\lambda_J$ increases, the influence of topological edge localization mitigates this error, yielding results more consistent with the exact dynamics. This observation leads us to favour a larger step size in our study. To quantify this, we calculated the density-dependent radial distribution, denoted as $R_{n}^{2}(t) = \sum_{i}in_{i}(t)$, and depicted it in Fig.~\ref{fig:radial}. For $\lambda_J = 0.0$, the IBM machine result (square dashed line) significantly diverges from the exact result (square solid line) over time. However, as the values of $\lambda_J$ increase, the discrepancy diminishes over numerous steps. This highlights the importance of localization effects in compensating for trotter, gate, and inherent system errors, thus enhancing the reliability of our QW simulations.
\begin{figure*}[!b]
  \centering
  \begin{minipage}[b]{0.48\columnwidth}
    \includegraphics[width=1\columnwidth]{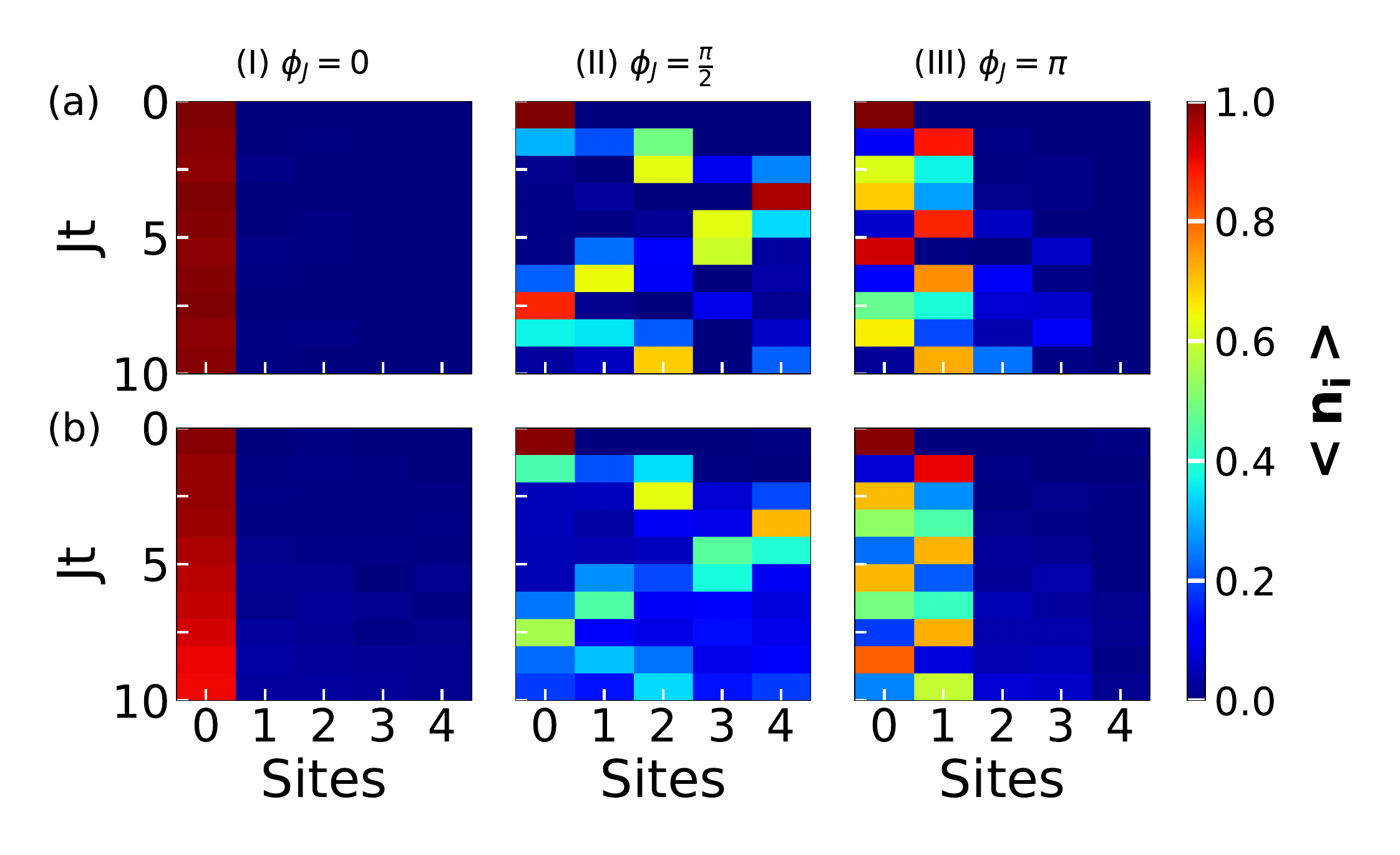}
    \caption{The figure illustrates the density evolution of quantum walkers initiated with the state $\ket{\psi}_I = \hat{c}^\dagger_0 \ket{vac}$. Panels (a) and (b) present results from the "Q-sim" and the readout error mitigated (REM) results from "ibm\_oslo", respectively. Here we consider $\lambda_J = 0.9$ and and $L=5$.}
    \label{fig:left_edge}
  \end{minipage}
  \begin{minipage}[b]{0.48\columnwidth}
    \includegraphics[width=1\columnwidth]{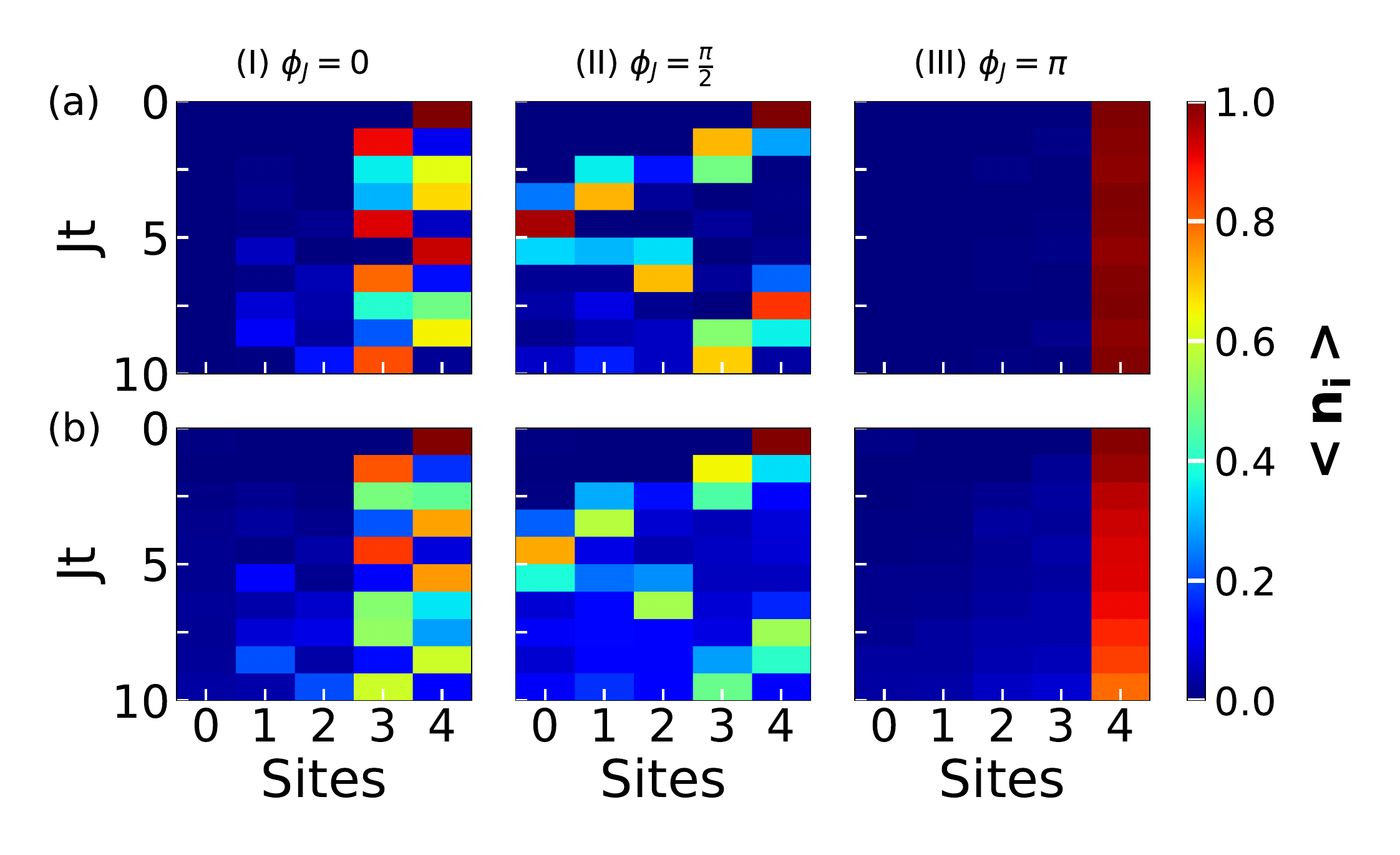}
    \caption{The figure illustrates the density evolution of quantum walkers initiated with the state $\ket{\psi}_{II} = \hat{c}^\dagger_4 \ket{vac}$. Panels (a) and (b) present results from "Q-sim" and the readout error mitigated (REM) results from
    "ibm\_oslo", respectively. Here we consider $\lambda_J = 0.9$ and and $L=5$.}
    \label{fig:right_edge}
  \end{minipage}
\end{figure*}

\begin{figure}[!t]
    \centering
    \includegraphics[width =0.7\columnwidth]{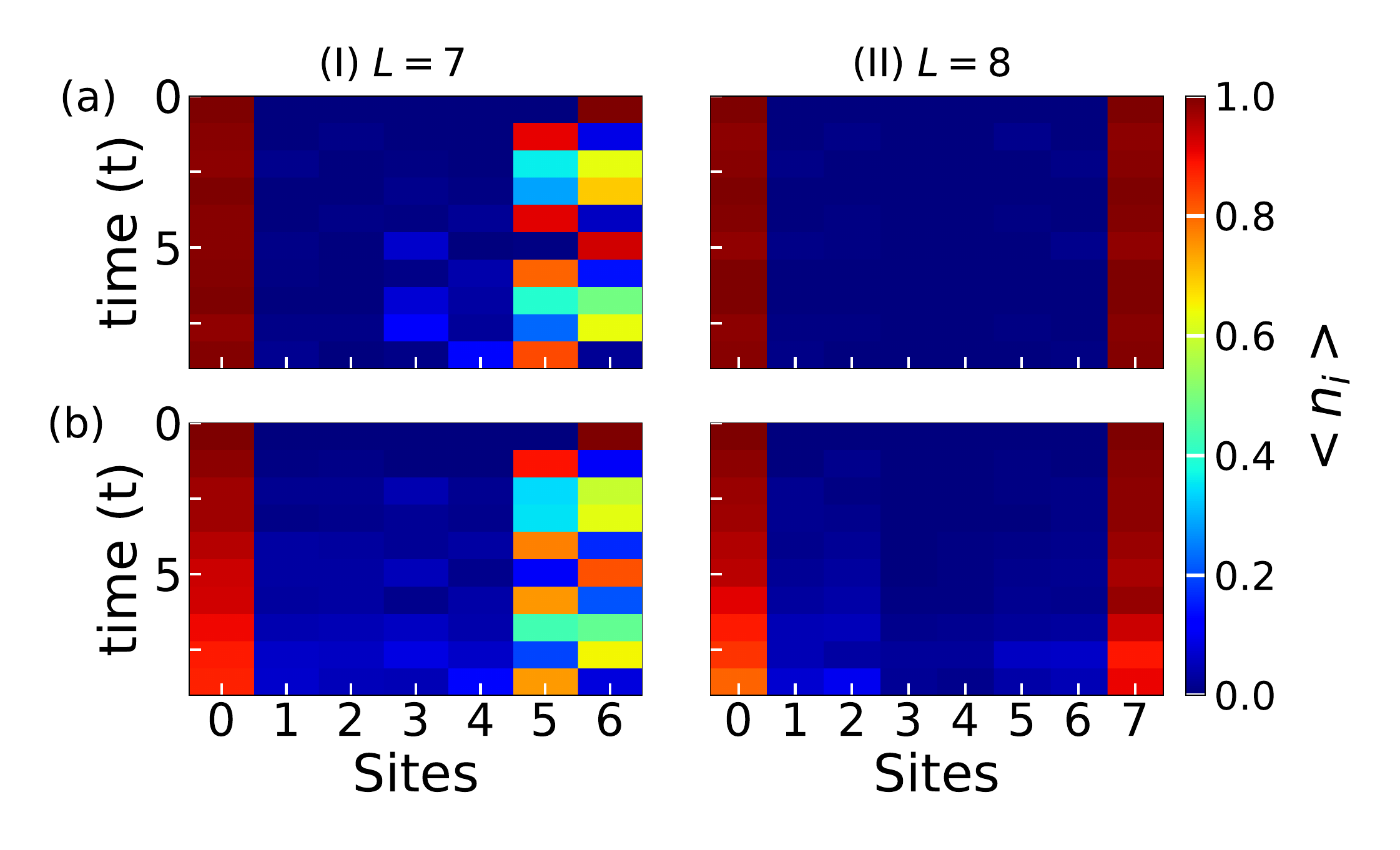}
    \caption{Figures (I) and (II) depicts the density evolution of quantum walkers initialized with the state $\ket{\psi} = \hat{c}^\dagger_0 \hat{c}^\dagger_6\ket{vac}$ and $\ket{\psi} = \hat{c}^\dagger_0 \hat{c}^\dagger_7\ket{vac}$, respectively. Panels (a) and (b) showcase results from "Q-sim" and the readout error mitigated (REM) results from "ibm\_brisbane," respectively. Here, we consider $\lambda_J = 0.9$ and $\phi_J=0$.}
    \label{fig:78-edge_state}
\end{figure}

We also analyze the role of the phase factor in edge dynamics with system size, $L=5$, utilizing the 7-qubit IBM machine "ibm\_oslo". For the case of $\phi=0$, the walker corresponding to initial state$|\psi_{I}\rangle$ localized at the left-edge of the lattice (see Fig.~\ref{fig:left_edge}(I)), but the walker corresponding to initial state $|\psi_{II}\rangle$ is not localized at the right-end of the lattice (see Fig.~\ref{fig:right_edge}(I)). The situation is reversed for $\phi=\pi$ as can be seen from the Fig.~\ref{fig:left_edge}(III) and Fig.~\ref{fig:right_edge}(III). The shifting of the edge state from one end to the other at a $\pi$ phase is due to the alteration of the strong and weak coupling between the bonds. For $0<\phi<\pi$, the edge state cannot be observed in the system because the topological effect vanishes for both initial states (see Fig.~\ref{fig:left_edge}(II) and Fig.~\ref{fig:right_edge}(II)). It is important to note that an edge state appears in the system due to the odd number of lattice sites considered. When an even number of lattice sites is used, two edge states are observed (see Fig.~\ref{fig:78-edge_state}). This lattice even–odd effect, which controls edge states in this model, will be discussed in the next section on two-particle dynamics. Thus, the phase factor's sensitivity is critical in topological quantum systems, making it important for various quantum information applications.
\begin{figure*}[!t]
  \centering
  \begin{minipage}[b]{0.5\columnwidth}
    \includegraphics[width=1.0\columnwidth]{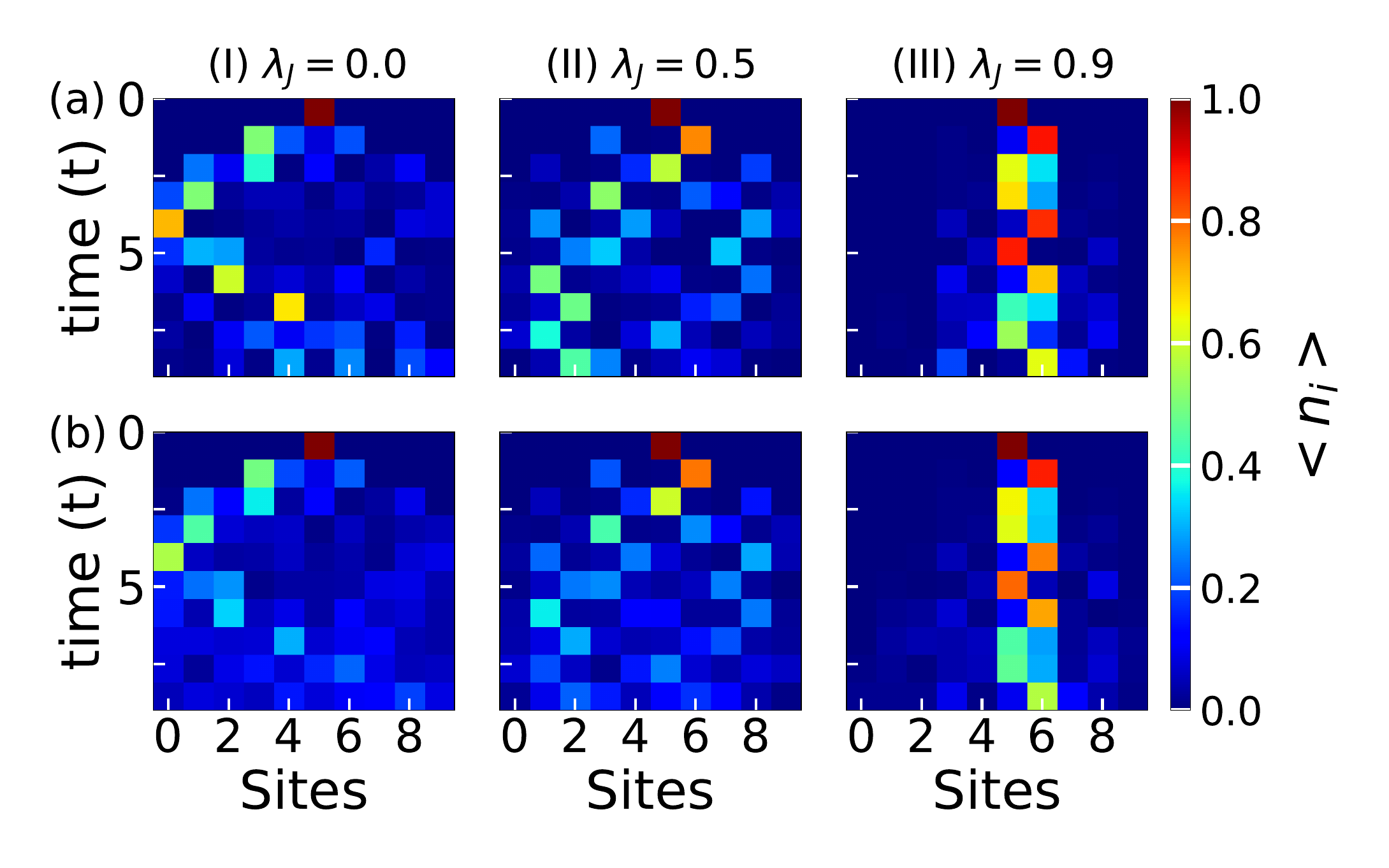}
    \caption{The figure displays the density evolution of quantum walkers initiated with the state $\ket{\psi}_{III} = \hat{c}^\dagger_5 \ket{vac}$. Panels (a) and (b) depict results from the "Q-sim" and the readout error mitigated (REM) results from "ibm\_brisbane," respectively. Here we consider $\phi_J = 0$ and and $L=10$.}
    \label{fig:bulk_QW}
  \end{minipage}
  \hfill
  \begin{minipage}[b]{0.47\columnwidth}
    \includegraphics[width=1\columnwidth]{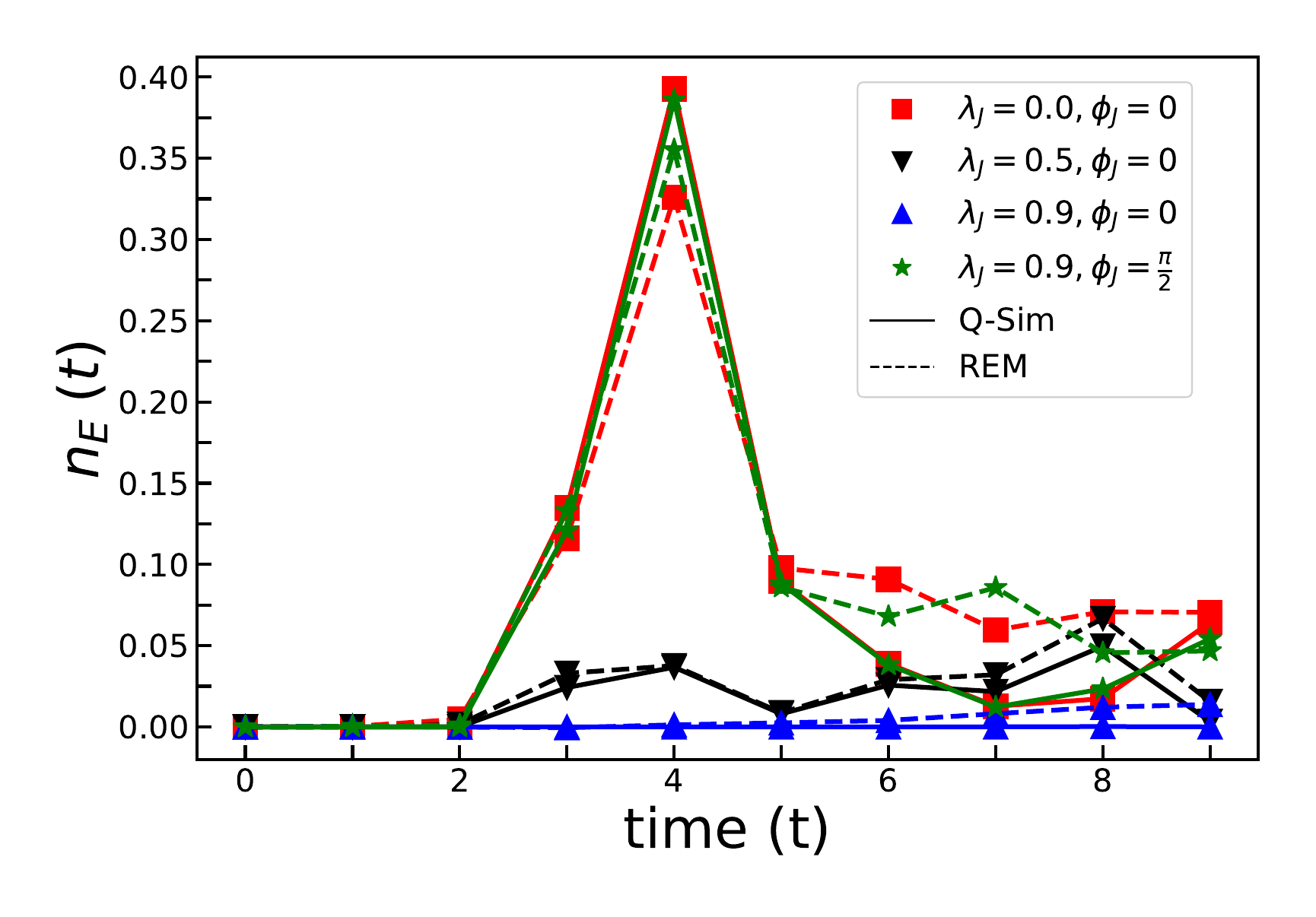}
    \caption{The illustrates $n_{E}(t)$ over time, $t$, with different $\lambda_J$ and $\phi_J$ values. The data includes results from "Q-sim" and the readout error mitigated (REM) results from "ibm\_brisbane".  Here we consider $L=10$.}
    \label{fig:ne_plot}
  \end{minipage}
\end{figure*}

Next, we explore the QW initiated from the bulk of the lattice with system size, $L = 10$. In Fig.~\ref{fig:bulk_QW}, the density evolution of the initial state defined by Eq.~\ref{eq:psi_2} for hopping modulation strengths $\lambda_J = 0.0, 0.5,$ and $0.9$, with the phase factor set to $\phi_J = 0$. In all cases, we observe that no localization phenomena occur in the QW. For $\lambda_J = 0.0$, the walker exhibits spreading of the wavefunction with a unidirectional bias owing to the basic trotterization. With an increase in $\lambda_J$, the walker's spreading is progressively suppressed as depicted in Fig.~\ref{fig:bulk_QW}(II) and (III) for $\lambda_J=0.5$ and $0.9$, respectively. If we carefully examine the Fig.~\ref{fig:bulk_QW}(I) and (II), the population density of the walker at the lattice edges gradually decreases with an increase in the hopping modulation strength, ultimately vanishing at $\lambda_J=0.9$. This behaviour is quantified by plotting the edge density, $n_{E}(t) = \frac{n_0(t) + n_9(t)}{2}$ as a function of time, $t$, as depicted in Fig.~\ref{fig:ne_plot}. For $\lambda_J=0$, the walker reaches the boundary, as evidenced by the peak in $n_{E}$ ( red square lines in Fig.~\ref{fig:ne_plot}). For $\lambda_J=0.5$, the density at the edges suppressed drastically ( down triangle black lines in Fig.~\ref{fig:ne_plot}) and tends to zero for $\lambda_J=0.9$ (up triangular blue lines in Fig.~\ref{fig:ne_plot}). This clearly indicates that for $\lambda_J \ne 0$, the walker fails to reach the lattice boundaries, suggesting an intriguing repulsion effect due to the presence of topologically protected edge states. This repulsion effect can be further explained within the context of the energy spectrum~\cite{sdas2013}, where the edge states are zero-energy states isolated from the continuum or scattering states. Consequently, particles initiating QWs from the bulk cannot access the edges, and vice versa, due to this isolation. When the phase factor $0<\phi_J<\pi$, the topological effect vanishes even for $\lambda_J =0.9$ (see Fig.~\ref{fig:left_edge}(II) and Fig.~\ref{fig:right_edge}(II)), thereby walker able to reach the lattice edges (similar to case of $\lambda_J =0$) without any repulsion effect from the edge states (see green star lines in Fig.~\ref{fig:ne_plot}). 


\subsection{Two-particle QW} 
In this section, we investigate the QW of two particles, both with and without interactions. We begin by studying the QW of two non-interacting particles placed at opposite ends of lattices with odd and even numbers of sites to understand the odd-even effect on the edge state. To focus on the edge states only, we set the phase factor $\phi_J = 0$ and the hopping modulation strength $\lambda_J = 0.9$. The dynamics reveal the emergence of a single edge state for $L=7$ and two edge states for $L=8$, as shown in Fig.~\ref{fig:78-edge_state}(I) and (II) respectively. The number of edge states depends on the odd-even parity of the system, which is characterized by the topological properties of the commensurate off-diagonal AAH model. For an even numbered lattice, this topological properties is characterized by the quantized Berry phase $\pi$ or or the emergence of a pair of degenerate edge states~\cite{PhysRevB.91.041402}. In an odd numbered lattice system, edge states are not degenerate anymore due to the breaking of the chiral or sub-lattice symmetry~\cite{sdas2013}. As a result a single edge state always appear on either edges of the lattice depending upon the phase factor,$\phi_J=0$ or $\pi$ due to the alteration of the strong and weak coupling between the bonds(see Fig.~\ref{fig:left_edge} and Fig.~\ref{fig:right_edge}). This phenomenon does not occur in lattices with an even number of sites, where two edge states always appear for phase factors,$\phi_J=0$ and $\pi$. These insights will drive the development of robust quantum algorithms and protocols, paving the way for more reliable and stable quantum computing technologies~\cite{Sarma2015}.

\begin{figure}[h!]
    \centering
    \includegraphics[width =0.7\columnwidth]{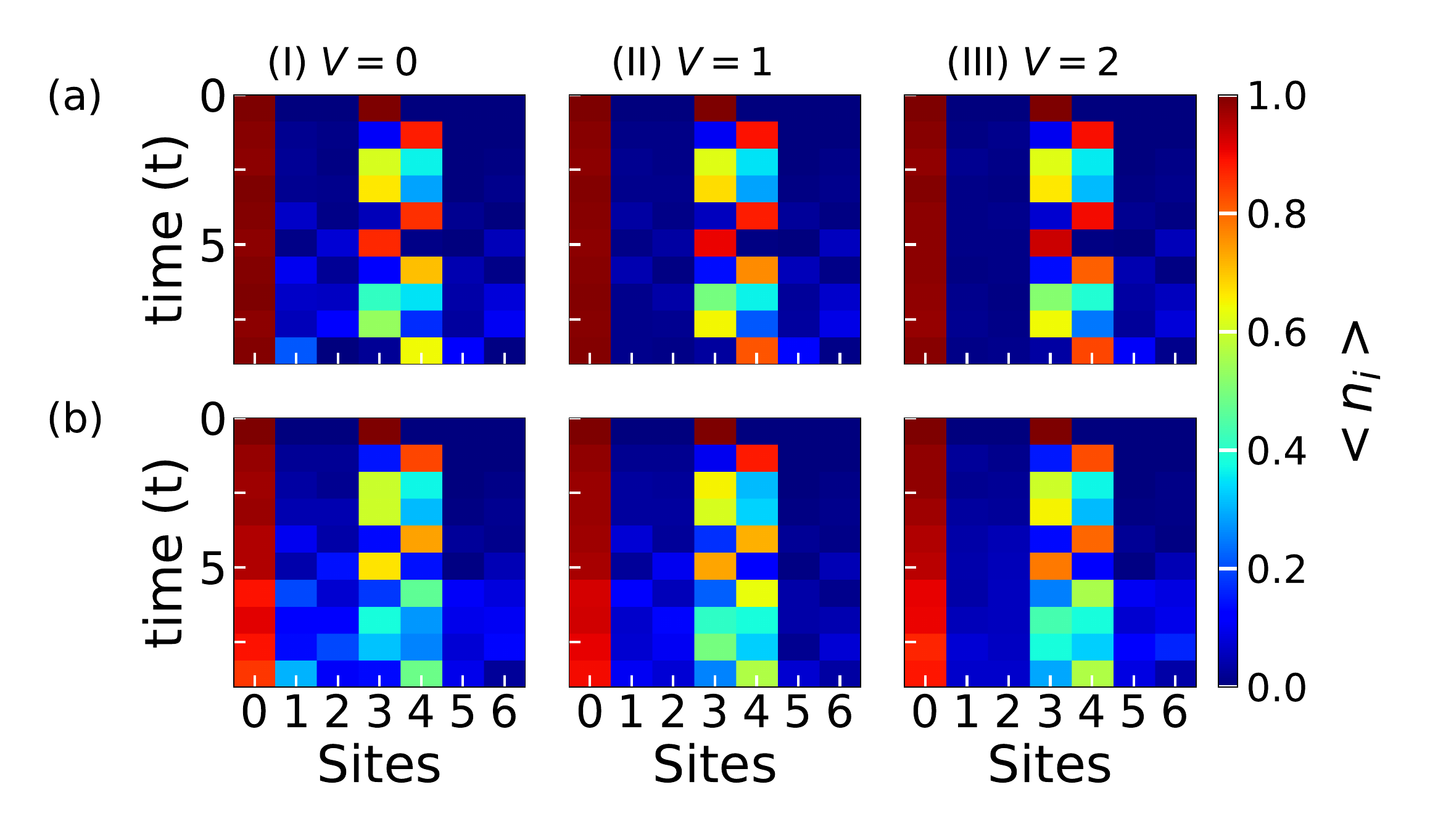}
    \caption{The figure depicts the density evolution of quantum walkers initialized with the state $\ket{\psi} = \hat{c}^\dagger_0 \hat{c}^\dagger_3 \ket{vac}$. Panels (a) and (b) showcase results from "Q-sim" and the readout error mitigated (REM) results from "ibm\_brisbane", respectively. Figures (I), (II), and (III) correspond to $V = 0$, $1$, and $2$, respectively. Here, we consider $L=7$, $\lambda_J=0.9$ and $\phi_J=0$.}
    \label{fig:03_state}
\end{figure}
Next, we investigate the effect of NN interaction in the QW with one walker starting from the edge and another from the bulk of the lattice, represented by the initial state, $|\psi\rangle = a_0 a_3 |0\rangle$. Fig.~\ref{fig:03_state} shows the density evolution for different interaction strengths. In this setup, we set the hopping modulation strength to $\lambda_J = 0.9$, the phase factor to $\phi_J=0$ and the lattice size to $L=7$ to obtain a single edge state. In the absence of interaction ($V=0$), the walker starting from the bulk able interact with the edge walker (see Fig.~\ref{fig:03_state}(I)). but on increase in the interaction strength (\(V=1\)), the bulk walker experiences reflection from the edge walker, as depicted in Fig.~\ref{fig:03_state}(II). Further increasing the interaction strength ($V=2$), the bulk walker is unable to approach the edge walker, displaying a distinct repulsion effect from the edge state due to the presence of interaction (see Fig.~\ref{fig:03_state}(III)). This repulsion effect has significant implications for quantum information dynamics, providing precise control over the transmission of two independent information. Tuning the interaction strength allows interference-free quantum communication, enabling the transmission of information along distinct paths without compromising signal integrity. This advancement has the potential to enhance controlled quantum communication protocols, contributing to improved reliability and efficiency in quantum information processing.


\begin{figure}[!h]
    \centering
    \includegraphics[width =0.9\columnwidth]{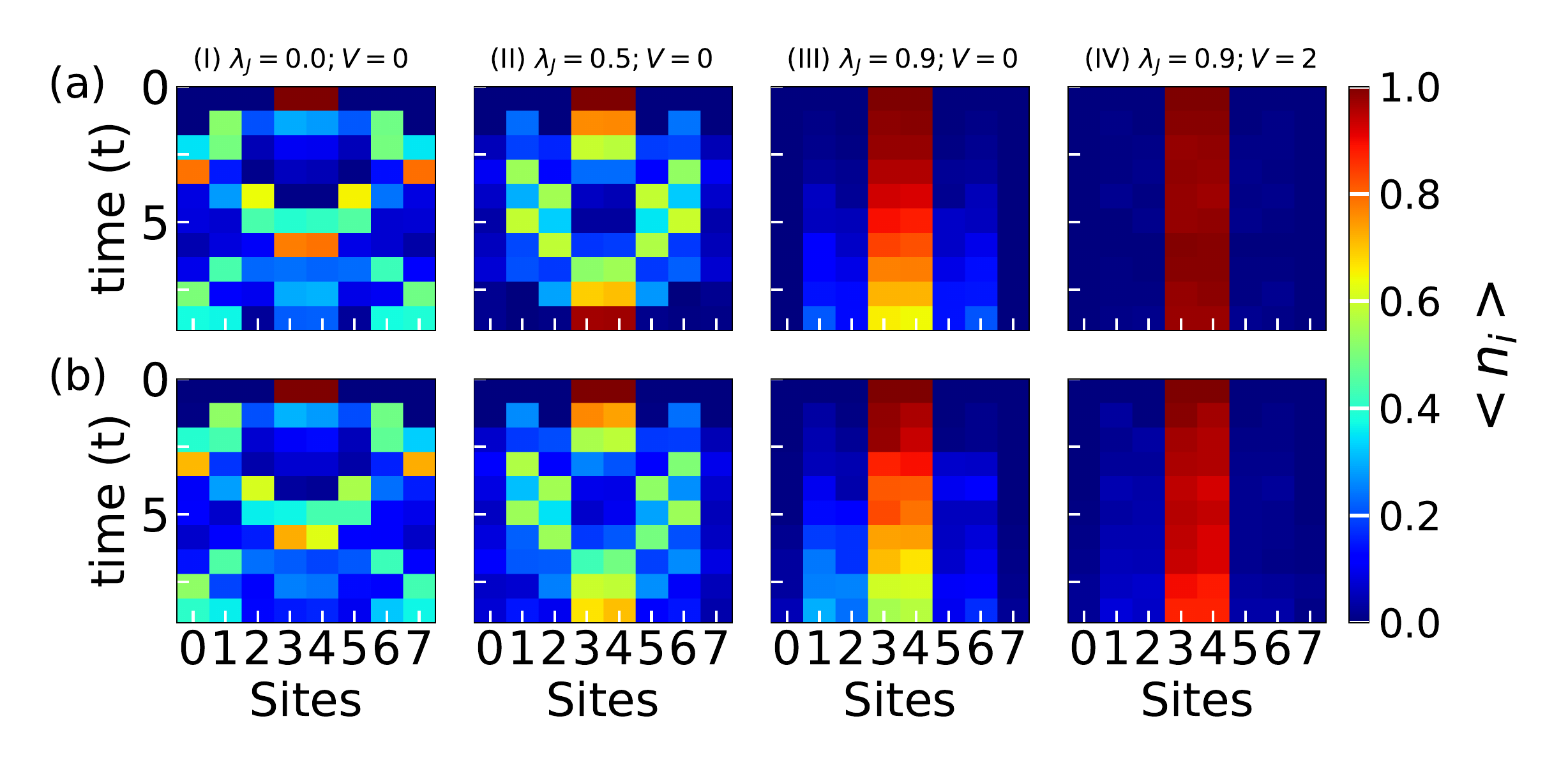}
    \caption{The figure shows the density evolution of quantum walkers initialized with the state $\ket{\psi} = \hat{c}^\dagger_3\hat{c}^\dagger_4 \ket{vac}$. Panels (a) and (b) showcase results from "Q-sim" and the readout error mitigated (REM) results from "ibm\_brisbane", respectively. Here, we consider $L=8$ and $\phi_J=0$.}
    \label{fig:34_state}
\end{figure}
\begin{figure}[!h]
    \centering
    \includegraphics[width =0.75\columnwidth]{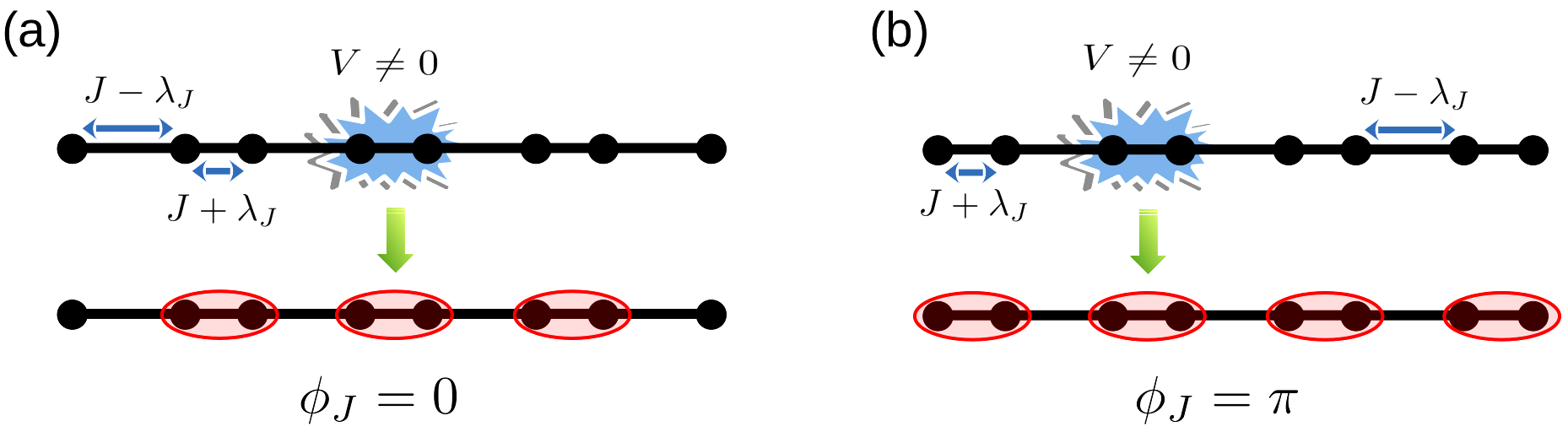}
    \caption{(a) and (b) correspond to the pictorial representation of the lattice for the phase factors $\phi_J=0$ and $\pi$, respectively. The red shaded region indicates the bonds where NN bound states are formed.}
    \label{fig:lattice_with_V}
\end{figure}

We also investigate the QW of two interacting particles situated within the lattice bulk, and the corresponding initial state is $|\psi\rangle = a_3 a_4|0\rangle$. Initially, we vary the hopping modulation strength while setting $V=0$ and $\phi_J=0$ , as shown in Fig.~\ref{fig:34_state}(I-III). For $\lambda_J=0$, the two walkers undergo independent particle QWs without affecting each others. Upon increasing $\lambda_J$, the walker spreading is gradually suppressed, and at $\lambda_J=0.9$, the two particles exhibit a collective behavior, moving as a composite particle with significantly slow spreading. 
When a weak interaction ($V$) is introduced, a complete localization or the formation of a two-particle NN bound state occurs in the bulk, as shown in Fig.~\ref{fig:34_state}(IV) for $V=2$. This is unusual because NN bound states typically form only at very strong NN interaction strengths~\cite{PhysRevLett.124.010404}. It is important to note that the formation of this bound state due to NN interaction depends on specific bonds between lattice sites, particularly where the hopping strength is minimal (see Fig.~\ref{fig:lattice_with_V}(a)); otherwise, it will not occur. Additionally, when $\phi_J \ne 0$, the bound state will not form for weak NN interaction (not shown), resulting in no localization phenomena in the dynamics except for $\phi_J=\pi$ due to the alteration of the strong and weak coupling between the bonds. 

\begin{figure}[t!]
    \centering
    \includegraphics[width =0.9\columnwidth]{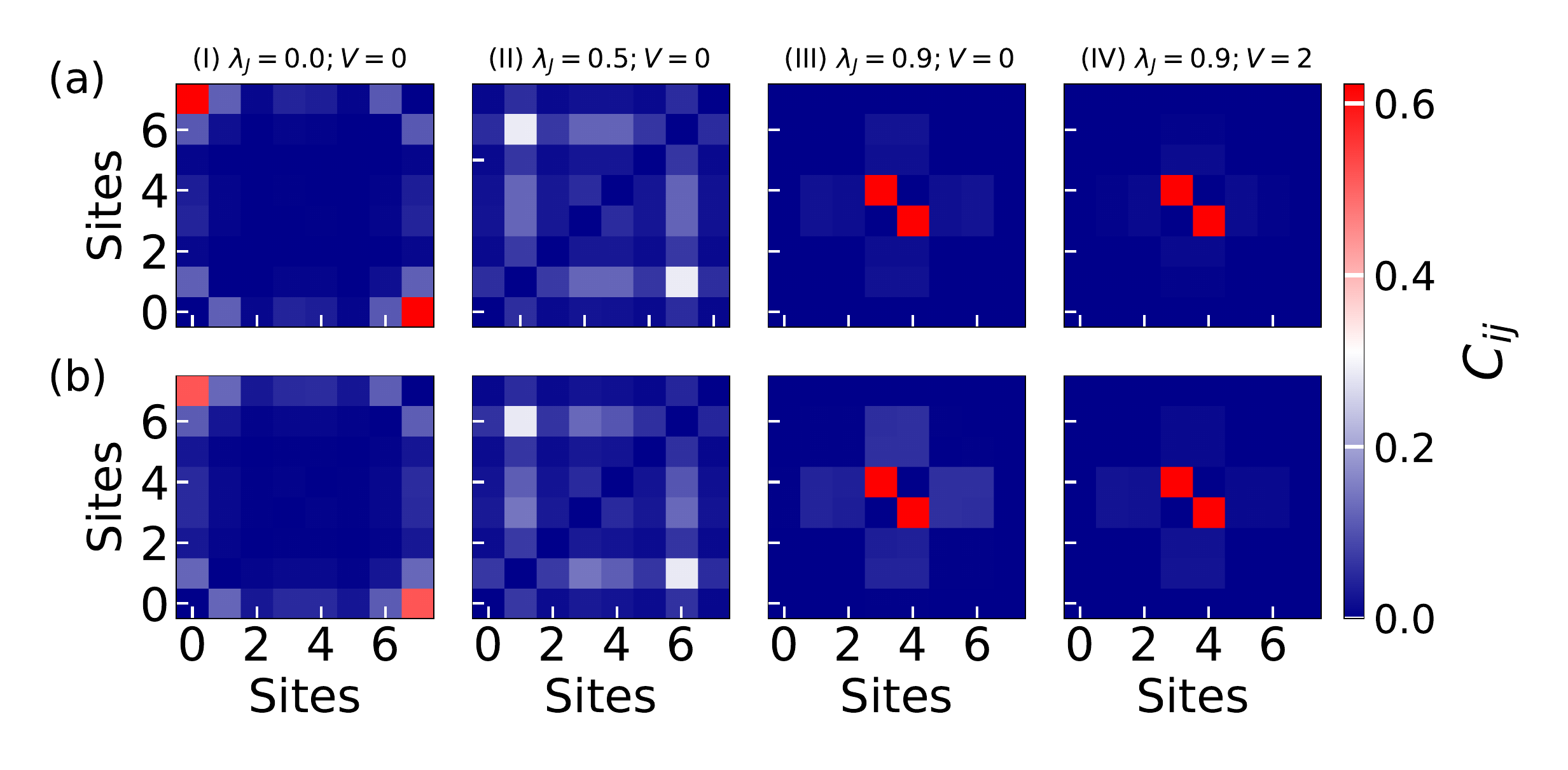}
    \caption{ The figure shows the two-particle quantum correlation, $C_{ij}$ for the initial state $\ket{\psi} = \hat{c}^\dagger_3\hat{c}^\dagger_4 \ket{vac}$ at time, $t=3J^{-1}$. Panels (a) and (b) showcase results from "Q-sim" and the readout error mitigated (REM) results from "ibm\_brisbane", respectively. Here, we consider $L=8$ and $\phi_J=0$.}
    \label{fig:34_state_corr}
\end{figure}

We further quantify this phenomena by calculating two-particle quantum correlation defined as-
\begin{equation}
    C_{ij}(t) = \braket{\psi(t)|\sigma^{z}_{i}|\psi(t)} \braket{\psi(t)|\sigma^{z}_{j}|\psi(t)},
\end{equation}
and plot it at time, $t=3J^{-1}$ in Fig.~\ref{fig:34_state_corr}. 
When $V=0$, two particles symmetrically spread from the center for $\lambda_J=0$, resulting in symmetric bright spots at the anti-diagonal edges on the correlation matrix, indicating fermionization of the particles (see Fig.~\ref{fig:34_state_corr}(I)). As $\lambda_J$ increases, the expansion of the walkers is suppressed, causing edge correlations to diminish and concentrate more at the center, signifying the co-walking of the two particles as can seen from Fig.~\ref{fig:34_state_corr}(II) and (III). At $V=2$ and $\lambda_J=0.9$, the two particles form a NN bound state, resulting in two bright patches on either side of the diagonal in the correlation plot (see Fig.~\ref{fig:34_state_corr}(IV)). This bound state formation in the interacting off-diagonal Aubry-Andr\'e-Harper model, facilitated by hopping modulation and weak NN interaction, has not been previously studied. In Panel-b of Fig.~\ref{fig:34_state_corr}, correlations calculated using IBM machine data show qualitative agreement with the Q-sim results.

\begin{figure}[!h]
    \centering
    \includegraphics[width =0.6\columnwidth]{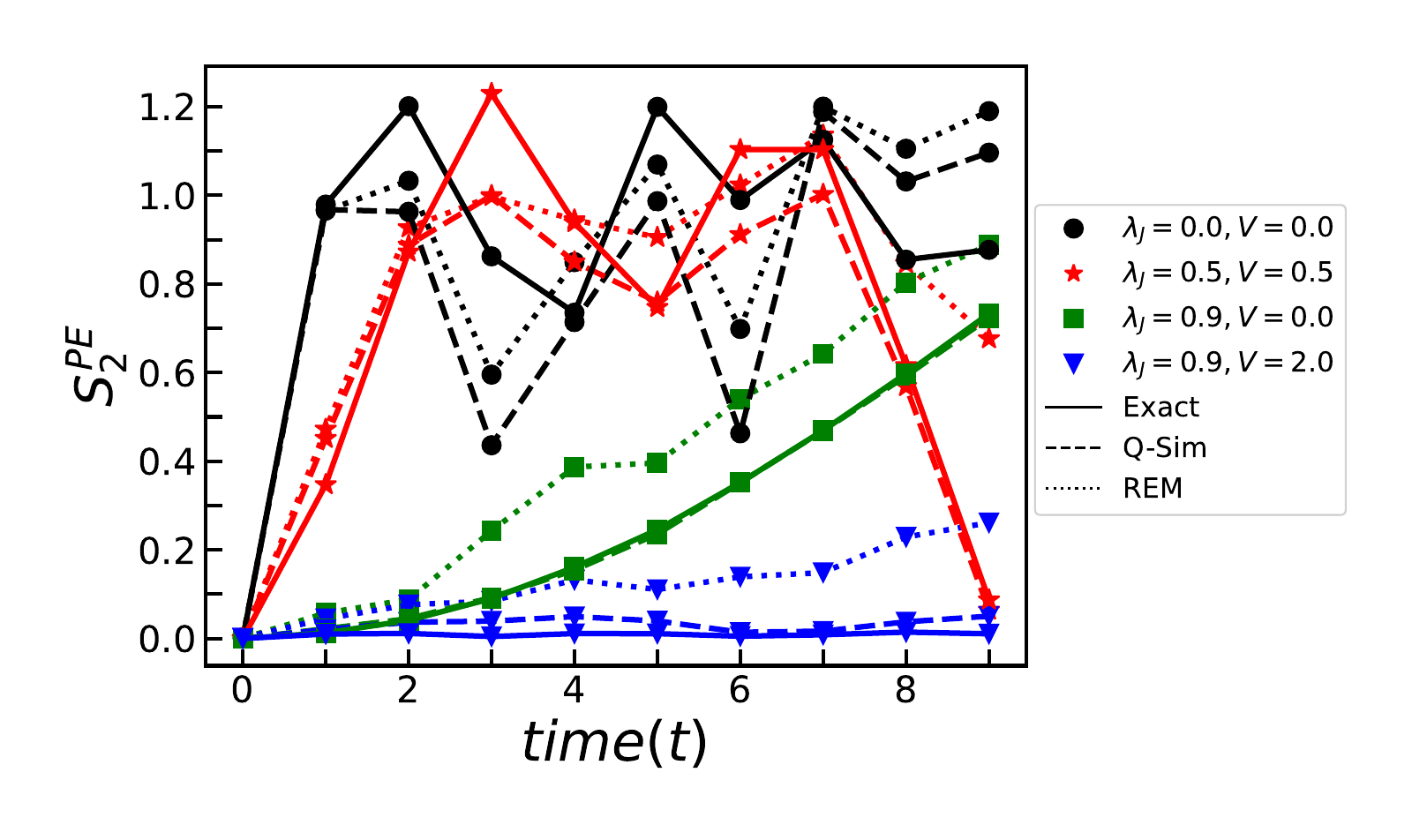}
    \caption{The figure depicts the participation entropy, $S_{2}^{PE}(t)$ as a function of time for the initial state $\ket{\psi} = \hat{c}^\dagger_3 \hat{c}^\dagger_4 \ket{vac}$ corresponding on-site density evolution shown in Fig.~\ref{fig:34_state}. Here, we consider $L=8$ and $\phi_J=0$.}
    \label{fig:entropy}
\end{figure}
For more clarity, we further analyze this bulk localization in the dynamics by examining the participation entropy as a function of time~\cite{Li2023} and  and is defined as -
\begin{equation}
\centering
    S_{k}^{PE}(t) = \frac{1}{1-k} log(\frac{1}{N}\sum_{i=0}^{L-1} p_{i}(t)^{k}),
    \label{eq:entropy}
\end{equation}
where $p_{i}(t)$ is the probability of the walker at site $i$ at time $t$, $k$ is the order of entropy and $N$ is the total number of particles in the system. In our study, we focus only on the second-order $(k=2)$ participation entropy, i.e.,  $S_{2}^{PE}(t) = -log [\frac{1}{N}\sum_{i=0}^{L-1} p_{i}(t)^{2}]$ and plot it as a function of time for different $\lambda_J$ and $V$ values in Fig.~\ref{fig:entropy}. For $V=0$ and $\lambda_J =0$ and $0.5$, the participation entropy initially shows rapid relaxation as the particles spread out. the participation entropy initially shows rapid relaxation as the particles spread out. Upon reflecting from the boundary, the particles converge, causing the entropy to decrease. This results in breathing-type oscillations in the dynamics, as shown by the black circle and red star lines in Fig.~\ref{fig:entropy}. When $\lambda_J=0.9$, the participation entropy is suppressed significantly (green square lines) reflecting a localization phenomena in the system. Next on introducing $V$, the two particles forms a NN bound state, as a result participation entropy shows almost a flat line indicating there are no spreading in the bulk of the system. But in the NISQ IBM machine, we can see some spreading in the dynamics due to many effects from inherent qubit noise to trotter error, still one can qualitatively study this off-diagonal AAH model due to its robust nature for $\lambda_J=0.9$. The observed phenomena of bulk localization in the presence of interaction bear substantial importance in the study of quantum many-body dynamics, quantum computation, and quantum information dynamics. 


\section{Conclusions}
We have systematically studied the QW in the off-diagonal AAH model with periodic hopping modulation (period $T=2$) using digital quantum computer. By considering different initial state, we have analyzed the impact of hopping modulation, phase factors and interaction on the QW of single and two particles. We have started QW with a particle placed at the lattice edge, and the hopping modulation strength is systematically varied. We observed that the edge state becomes apparent when the hopping modulation is strong enough. The robustness of this edge state becomes evident as it exhibits resilience against factors such as inherent quantum hardware noise and trotter error. This resilience is quantified through the analysis of the density-dependent radial distribution, and its reliability is further confirmed by comparisons with exact calculations. Furthermore, we have studied the impact of the phase factor on edge dynamics and topological repulsion effect from the edge state in the QW under sufficiently strong hopping modulation. We also have studied the interplay of NN interaction and topology, where the QW exhibit a repulsion effect from the topological edge state. This phenomena suggests potential application of the QW for interference-free quantum communication. Additionally, we have studied the QW of two interacting particles starting from the bulk of the lattice, revealing the formation of bound states at weak interaction. We have benchmarked all our results using the IBM quantum computer, which has not yet been fully explored. Our benchmark digital quantum simulation results open up opportunities to study new and complex condensed matter phenomena, such as dynamical phase transitions~\cite{Jacek_DPT,Heyl_2018}, quantum many-body scars~\cite{Turner2018}, and time crystals~\cite{Sacha_2018}, which are difficult to realize on other platforms.

Recent technical advancements in digital quantum simulation, including quantum circuit optimization~\cite{Nam2018,PhysRevA.98.032309}, quantum error correction~\cite{QEC1,PhysRevA.101.022305}, randomized benchmarking~\cite{PhysRevLett.112.240504,PhysRevA.99.052323}, and sophisticated error mitigation techniques~\cite{Li2019,PhysRevLett.122.180501,Kandala2019,PRXQuantum.2.040330}, have enabled the study of more complex systems on a larger scale. These advancements improve the precision and practicality of simulations, allowing for the exploration of complex quantum phenomena and extending the limits of quantum computing. Harnessing these advancements offers the opportunity to gain profound insights into the essential properties of quantum systems and their potential applications in quantum technologies.

\section{Acknowledgment}
We acknowledge the use of IBM Quantum services for this work. The views expressed are those of the authors, and do not reflect the official policy or position of IBM or the IBM Quantum team. MKG acknowledge fruitful discussions with Bhanu Pratap Das, Beno\^{i}t Vermersch, Tapan Mishra, Aniket Rath, Vittorio Vitale and Sudeshna Madhual. MKG acknowledges the support received for this research, which was partially funded by the National Science and Technology Council (NSTC) of Taiwan through NSTC 112-2811-M-007-050.
\begin{appendices}

\section{Jordan-Wigner Transformation}
\label{A:JW}

The Jordan-Wigner transformation~\cite{Jordan1928} is a mathematical technique in condensed matter physics that exactly maps one dimensional fermions to hardcore bosons or spin-1/2 magnetic moments or vice versa. In quantum computing, this transformation is employed to map fermionic operators to qubits, enabling the simulation of fermionic systems on quantum computers~\cite{PhysRevB.104.035118,BRAVYI2002210}. For a fermionic system, the creation and annihilation operators can be represented in terms of Pauli matrices, which are fundamental components in quantum computation. The Pauli matrices are defined as-
\begin{equation}
    \begin{split}
        \sigma_x &= \ket{0}\bra{1} + \ket{1}\bra{0} \\
        \sigma_y &= -i\ket{0}\bra{1} + i\ket{1}\bra{0} \\
        \sigma_z &= \ket{0}\bra{0} - \ket{1}\bra{1} 
    \end{split}
\end{equation}
where $\ket{0}$ and $\ket{1}$ represent the two orthogonal basis states of a spin-1/2 system.
Now the creation and annihilation operators are defined as:
\begin{equation}
    \begin{split}
        \hat{c}^\dagger = \frac{\sigma_x -i\sigma_y}{2} = \sigma^{-}\;\;\; ; \;\;\;
        \hat{c} = \frac{\sigma_x +i\sigma_y}{2} = \sigma^{+}
    \end{split}
\end{equation}
Using these definitions, we can express the spin operators as linear combinations of the creation and annihilation operators:
\begin{equation}
    \begin{split}
        S_x = \frac{\hat{c}^\dagger + \hat{c}}{2}; \;\;\;\;\;\;
        S_y = \frac{\hat{c}^\dagger - \hat{c}}{2i} ;\;\;\;\;\;\;
        S_z = \frac{\hat{c}^\dagger\hat{c} - \hat{c}\hat{c}^\dagger}{2}
    \end{split}
\end{equation}
where $S_x$, $S_y$, and $S_z$ are the Pauli spin operators.

But we also have to incorporate the anti-commutation relation of fermionic operators during the mapping and it is done by interspersing Z operators into the construction of the qubit operator which  emulates the correct anti-commutation. For a 1D lattice with N sites, the corresponding transformation can be written as, 
\begin{equation}
    \begin{split}
        \hat{c}_0^\dagger &=  \hat{\sigma}^{-}_{0}\otimes I_1 \otimes I_2 \dots I_N \\
        \hat{c}_1^\dagger &= {\sigma_{z}}_{0} \otimes \hat{\sigma}^{-}_{1}\otimes I_2\dots I_N\\
        \vdots \\
        {\hat{c}_N}^\dagger &= {\sigma_{z}}_{0} \otimes {\sigma_{z}}_{1} \otimes {\sigma_{z}}_{2}\dots \otimes{\hat{\sigma}^{-}}_{N}
    \end{split}
\quad \quad \quad \text{and}\quad \quad \quad
    \begin{split}
        \hat{c}_0 &=  \hat{\sigma}^{+}_{0}\otimes I_1 
 \otimes I_2 ...... I_N \\
        {\hat{c}}_1 &= {{\sigma}_z}_0 \otimes \hat{\sigma}^{+}_{1}\otimes I_2\dots I_N\\
        \vdots \\
        {\hat{c}}_N &= {\sigma_{z}}_{0} \otimes {\sigma_z}_1 \otimes {\sigma_z}_2\dots \otimes\hat{\sigma}^{+}_{N}.
    \end{split}
\end{equation}
This transformation is known as Jordan-Wigner Transformation and $\otimes$ denotes the tensor product.
Now any Hamiltonian under this transformation can be written as,
\begin{equation}
    H = \sum_{i} h_j P_j = \sum_{i} h_j \Pi_{i}\sigma_i^j. 
\end{equation}
Where $h_j$ is the scalar coefficient and $P_j = \Pi_{i}\sigma_i^j$ ,where $\sigma_i^j $ is the Pauli matrices.

Here we have given a simple example of  the Jordon-Wigner transformation. We consider a Bose-Hubbard Hamiltonian for a single particle in a one-dimensional lattice with $L=3$ lattice sites can be represented in terms of second quantized fermionic creation and annihilation operators.

\begin{equation}
    H= -J(\hat{c}_0^\dagger \hat{c}_1 + \hat{c}_1^\dagger \hat{c}_0+ \hat{c}_1^\dagger \hat{c}_2+ \hat{c}_2^\dagger \hat{c}_1)
    \label{eq:B-hamiltonian}
\end{equation}

Now using Jordan Wigner transformation, we can rewrite the above Hamiltonian in terms of the product of Pauli matrices. The corresponding mapping is given below,
\begin{equation}
    \begin{split}
        \hat{c}_0^\dagger &=  \hat{\sigma}^{-}_{0}\otimes I_1 
 \otimes I_2 \\
        {\hat{c}_1}^\dagger &= {{\sigma}_z}_0 \otimes \hat{\sigma}^{-}_{1}\otimes I_2\\
        {\hat{c}_2}^\dagger &= {{\sigma}_z}_0 \otimes {{\sigma}_z}_1 \otimes \hat{\sigma}^{-}_{2}
    \end{split}
\quad \quad \quad \text{and}\quad \quad \quad
    \begin{split}
        \hat{c}_0 &=  \hat{\sigma}^{+}_{0}\otimes I_1 
 \otimes I_2 \\
        {\hat{c}}_1 &= {\sigma_z}_0 \otimes \hat{\sigma}^{+}_{1}\otimes I_2\\
        {\hat{c}}_2 &= {\sigma_z}_0 \otimes {\sigma_z}_1 \otimes \hat{\sigma}^{+}_{2}.
    \end{split}
\end{equation}
Now using the above relation and calculate the terms below as -
\begin{equation}
    \begin{split}
    \hat{c}_0^\dagger \hat{c}_1 + \hat{c}_1^\dagger \hat{c}_0&=(\hat{\sigma}^{-}_{0}\otimes I_1\otimes I_2)({\sigma_z}_0 \otimes \hat{\sigma}^{+}_{1}\otimes I_2) +  ({\sigma_z}_0 \otimes \hat{\sigma}^{-}_{1}\otimes I_2)( \hat{\sigma}^{+}_{0}\otimes I_1 
 \otimes I_2)\\
 &=(\hat{\sigma}^{-}_{0} {\sigma_z}_0  \otimes \hat{\sigma}^{+}_{1} \otimes I_2) + ({\sigma_z}_0 \hat{\sigma}^{+}_{0} \otimes \hat{\sigma}^{-}_{1}\otimes I_2) \quad \text{[Using the relation: $(A \otimes B).(C \otimes D) = (A.C) \otimes (B.D)$]}\\
&=(\frac{{\sigma_x}_{0} -i{\sigma_y}_{0}}{2}) {\sigma_z}_0 \otimes (\frac{{\sigma_x}_{1} +i{\sigma_y}_{1}}{2}) \otimes I_2 +  {\sigma_z}_0 (\frac{{\sigma_x}_{0} +i{\sigma_y}_{0}}{2})\otimes(\frac{{\sigma_x}_{1} -i{\sigma_y}_{1}}{2}) \otimes I_2\\
&=(\frac{{\sigma_x}_{0} -i{\sigma_y}_{0}}{2})  \otimes (\frac{{\sigma_x}_{1} +i{\sigma_y}_{1}}{2}) \otimes I_2 +  (\frac{{\sigma_x}_{0} +i{\sigma_y}_{0}}{2})\otimes(\frac{{\sigma_x}_{1} -i{\sigma_y}_{1}}{2}) \otimes I_2\\
&=\frac{{\sigma_x}_{0} \otimes {\sigma_x}_{1}\otimes I_2}{2} + \frac{{\sigma_y}_{0} \otimes {\sigma_y}_{1}\otimes I_2}{2} 
    \end{split}
\end{equation}
Similarly, we get,
\begin{align*}
 {\hat{c}_1}^\dagger {\hat{c}}_2 + {\hat{c}_2}^\dagger {\hat{c}}_1=\frac{  I_0 \otimes {\sigma_x}_{1} \otimes {\sigma_x}_{2}}{2} + \frac{ I_0 \otimes {\sigma_y}_{1} \otimes {\sigma_y}_{2} }{2}    
\end{align*}
Thus the Bose-Hubbard hamiltonian after Jordan-Wigner mapping will look like,
\begin{equation}
\hat{H} = -\frac{J}{2} ({\sigma_x}_{0} \otimes {\sigma_x}_{1}\otimes I_2 + {\sigma_y}_{0} \otimes {\sigma_y}_{1}\otimes I_2 + I_0 \otimes {\sigma_x1}_{1} \otimes {\sigma_x}_{2} + I_0 \otimes {\sigma_y}_{1} \otimes {\sigma_y}_{2})
\end{equation}
We remove the $\otimes$ and Identity from the above equation and rewrite the above Hamiltonian in a simple form as shown below
\begin{equation}
    \hat{H} = -\frac{J}{2} \sum_{i=1}^{L-1}\sigma_x^i \sigma_x^{i+1} + \sigma_y^{i} \sigma_y^{i+1}
    \label{JW-BH}
\end{equation}
In this way, using Jordon-Wigner transformation fermionic operator can be easily mapped to Pauli spin operator, so that one can easily construct the quantum gates for quantum computations. But, in the case of bosonic systems, this transformation will not worked except for the hardcore limit, since in the hardcore limit Bose-Hubbard Hamiltonian equivalent to the spinless Fermi-Hubbard Hamiltonian. 
\end{appendices}

\begin{appendices}
\section{Suzuki-Trotter Decomposition}
\label{B:trotter}
The Suzuki-Trotter decomposition is a technique used in quantum computing to simulate the evolution of a quantum system~\cite{Hatano2005,Daley2022}. The basic idea behind the Suzuki-Trotter decomposition is to break up the time evolution operator of a quantum system into a sequence of simpler operators that can be more easily simulated on a quantum computer. The decomposition is based on the Baker-Campbell-Hausdorff (BCH) formula, which expresses the product of two non-commuting operators as a sum of commutators.
Consider a Hamiltonian containing only the local nearest neighbour interactions $h_i$'s can be written as
\begin{equation}
    H = \sum_{i=1}^L h_i= \sum_{i \in odd} h_i+ \sum_{i \in even } h_i = \hat{H}_{o} + \hat{H}_{e},
\end{equation}
where $\hat{H}_{e}$ and $\hat{H}_{o}$ do not commute with each other.
Then the exact time evolution operator becomes- 
\begin{equation}
    \hat{\mathcal{U}}(t)= e^{-{i\hat{H}t}}=e^{-{i(\hat{H}_{o} + \hat{H}_{e})t}}
\end{equation}
Upon using  the BCH formula the exact time evolution operator can be approximated as 
\begin{equation}
    \hat{\mathcal{U}}(t) \approx e^{-{i\hat{H}_{o}t}} e^{-{i\hat{H}_{e}t}} + \mathcal{O}(t^2).
\end{equation}
This is known as first-order decomposition and error in each step is the order of $\mathcal{O}(t^2)$.
For a long time of evolution, we will break the evolution time $t$ into $n$ number of time steps or trotter steps of $\delta t (= t/n) $ duration. Then, time evolution can be approximated using the first-order Trotter decomposition formula is given by -
\begin{equation}
    \hat{\mathcal{U}}(t) \approx (e^{-{i\hat{H}\delta t}})^{n} \approx (e^{-{i\hat{H}_{o}t/n}} e^{-{i\hat{H}_{e}t/n}})^n + \mathcal{O}(t^2/n).
\end{equation}
We can see that for $n \rightarrow \infty$ i.e for large number of trotter steps or small time interval step $\delta t$, the unitary evolution becomes more accurate as shown below- 
\begin{equation}
    \hat{\mathcal{U}}(t) \approx \lim_{n\to\infty} (e^{-{i\hat{H}_{o}t/n}} e^{-{i\hat{H}_{e}t/n}})^n
\end{equation}
The error in the first-order Suzuki-Trotter decomposition\cite{Nielsen} can be further reduced by considering higher-order decomposition. For second-order decomposition one has to rewrite the time evolution as -
\begin{equation}
    \hat{\mathcal{U}}(t) \approx e^{-{i\hat{H}_{o}t/2}} e^{-{i\hat{H}_{e}t}} e^{-{i\hat{H}_{o}t/2}} + \mathcal{O}(t^3),
\end{equation}
here the error associated with each step is the order of $\mathcal{O}(t^3)$. For $n$ step with step size of $\delta t = t/n$, the evolution operator becomes -
\begin{equation}
    \hat{\mathcal{U}}(t) \approx (e^{-{i\hat{H}_{o}\delta t/2}} e^{-{i\hat{H}_{e}\delta t}} e^{-{i\hat{H}_{o}\delta t/2}})^n + \mathcal{O}(t^3/n^2).
\end{equation}
The Suzuki-Trotter decomposition is widely used in condensed matter physics and quantum chemistry to simulate the behavior of complex quantum systems. However, it is important to note that the accuracy of the method depends on the order of the decomposition and the size of the time step, and may require some optimization for specific systems.
\end{appendices}
\bibliography{references}
\end{document}